\documentclass[twoside,12pt]{article}
\input{epsf.tex}
\usepackage[dvips]{graphics,epsfig}
\usepackage[latin1]{inputenc}
\usepackage{textcomp}
\def\laq{\raise 0.4 ex \hbox{$<$}\kern -0.8 em\lower 0.62 ex\hbox{$\sim$}}
\def\gaq{\raise 0.4 ex \hbox{$>$}\kern -0.7 em\lower 0.62 ex\hbox{$\sim$}}

\def\beq{\begin{equation}}
\def\eeq{\end{equation}}
\def\beqa{\begin{eqnarray}} 
\def\eeqa{\end{eqnarray}}

\begin{document}
\pagestyle{plain}

\begin{flushright}
DF/IST--8.2008\\
February 15, 2009
\end{flushright}
\vspace{15mm}

\begin{center}

{\Large\bf Stability Conditions For a Noncommutative Scalar Field Coupled to Gravity}

\vspace*{1.0cm}

Orfeu Bertolami$^{*}$ and Carlos A. D. Zarro$^{**}$\\
\vspace*{0.5cm}
{Instituto Superior T\'ecnico, Departamento de F\'\i sica, \\
Av. Rovisco Pais, 1049-001 Lisboa, Portugal}\\

\vspace*{2.0cm}
\end{center}

\begin{abstract}

\noindent
We consider a noncommutative scalar field with a covariantly constant noncommutative parameter in a curved space-time background.  For a potential as a noncommutative polynomial it is shown that the stability conditions are unaffected by the noncommutativity, a result that is valid irrespective whether  space-time has horizons or not.

\end{abstract}

\vfill
\noindent\underline{\hskip 140pt}\\[4pt]
{$^{*}$ Also at Instituto de Plasmas e Fus\~{a}o Nuclear, Instituto Superior T\'ecnico,
Lisboa} \\
{E-mail address: orfeu@cosmos.ist.utl.pt} \\
\noindent
{$^{**}$ Also at Instituto de Plasmas e Fus\~{a}o Nuclear, Instituto Superior T\'ecnico,
Lisboa} \\
{E-mail address: zarro@fisica.ist.utl.pt}

\newpage


\section{Introduction}
Noncommutative geometry is believed to be a fundamental ingredient of quantum gravity \cite{Connes:1996gi} and it is shown to arise, under conditions, in String Theory \cite{Seiberg:1999vs}. Noncommutativity introduces a minimum length scale and can be implemented by generalizing the Heisenberg-Weyl algebra of Quantum Mechanics \cite{Snyder:1947}. This scale is presumably associated to the Planck length $L_{P}$, so that the structure of the space-time is assumed to be altered at this scale. Given its potentialities, noncommutative features can be implemented in Quantum Field Theories (for reviews, see $e.g.$ Refs. \cite{Szabo:2001kg, Douglas:2001ba}), however, it is shown that the existence of a minimum length scale does not solve the problem of IR divergences and it actually introduces additional unitarity and causality problems. Other critical issues associated with noncommutative geometry involve the violation of translational invariance \cite{Bertolami:2003nm} and the question of noncommutative fields on a classical cosmological background \cite{Lizzi:2002ib,Bertolami:2002eq}. 

Another interesting subject of research associated to noncommutative geometry concerns extensions of the quantum mechanics Heisenberg-Weyl algebra in order to generalize quantum mechanics both at configuration space level as well as at full phase space \cite{Gamboa:2001fg,Zhang:2004yu,Bertolami:2005jw,Acatrinei:2003id,Bertolami:2005ud,Bastos:2006kj,Bastos:2006ps}. Furthermore, noncommutative quantum cosmological models in the context of the minisuperspace Kantowski-Sachs metric have also been studied \cite{GarciaCompean:2001wy,Barbosa:2004kp,Bastos:2007bg}. Phase space noncommutative extensions which exhibit momenta space noncommutativity  yield particularly interesting new feature in what concerns the selection of states for the early universe \cite{Bastos:2007bg}.

In this work we examine the stability of noncommutative scalar fields with a polynomial potential in a curved space-time background. For this purpose we consider extensions of the positive energy theorem for gravity as originally deduced in Ref. \cite{Witten:1981mf}. The positive energy theorem states that the total gravitational energy cannot be negative if matter fields satisfy the dominant energy condition \cite{Witten:1981mf,ChoquetBruhat:1985xy}. This establishes the classical and semi-classical stability of the Minkowski space-time. We consider the extension of this theorem that includes other kinds of fields such as scalar and vector fields. It is interesting that this setup allows, for instance,   obtaining Bogolmony bounds for electromagnetic fields \cite{Gibbons:1982fy}. The stability of supergravity gauge theories was examined in Ref. \cite{Gibbons:1983aq}, where it is shown the stability of supersymmetric theories in AdS spaces even when they exhibit  negative local energy densities. This method was generalized to tackle situations where fields that do not admit a supersymmetric extension in Ref.\cite{Boucher:1984yx} and used for studying the stability conditions for scalar fields non-minimally coupled to gravity \cite{Bertolami:1987wj}. The generalization of the positive energy theorem to include black hole-type  space-time was discussed in Ref. \cite{Gibbons:1982jg}.

In our approach, we shall obtain the stability conditions for noncommutative scalar fields in the presence of gravity using the method of Refs. \cite{Boucher:1984yx,Bertolami:1987wj}. This is achieved through a model in which the noncommutativity on scalar field is implemented via a  Moyal product adapted to a curved space-time with a covariantly constant noncommutative parameter \cite{Lizzi:2002ib} and an additional  condition to ensure the associativity of the noncommutative polynomial scalar field potential, as suggested in Ref. \cite{Bertolami:2002eq}. 

This Letter is organized as follows: in Section \ref{sec:model} we present our noncommutative scalar field model and the conditions for consistently coupling it to gravity. In Section \ref{sec:gpet} we discuss the commutative positive energy theorem and obtain its noncommutative counterpart.  In Section \ref{sec:stability} we obtain the stability conditions for noncommutative scalar fields with a  noncommutative polynomial potential. In Section \ref{sec:bhspt} we introduce space-times with horizons and show that the stability conditions previously obtained are also valid for these spaces. Finally Section \ref{sec:conclusions} contains our conclusions.


\section{The model}\label{sec:model}
In a Minkowski space-time  noncommutativity of fields is introduced via the so-called Moyal product \cite{Bayen:1977ha}

\beq \label{eq:notcovmoyal}
f \ast g=\sum_{n=0}^{\infty}\frac{(i/2)^{n}}{n!}\theta^{\alpha_{1}\beta_{1}}\cdots\theta^{\alpha_{n}\beta_{n}}\left(\partial_{\alpha_{1}}\cdots\partial_{\alpha_{n}}f\right)\left(\partial_{\beta_{1}}\cdots\partial_{\beta_{n}}g\right),
\eeq

\noindent where $\theta^{\mu\nu}$ is a constant noncommutative parameter. This parameter is related to the commutator between noncommutative coordinates in configuration space $[x^{\mu},x^{\nu}]=i\theta^{\mu\nu}$. 

This product is not covariant, thus when considering a curved space-time a natural implementation for a covariant Moyal product would involve instead $\theta^{\mu\nu}$ as a tensor and covariant derivatives \cite{Lizzi:2002ib, Bertolami:2002eq, BarcelosNeto:2002bh, Harikumar:2006xf}\footnote{Where f and g are in general tensor fields although their indices are omitted for simplicity.}

\beq \label{eq:covmoyal}
f \star g=\sum_{n=0}^{\infty}\frac{(i/2)^{n}}{n!}\theta^{\alpha_{1}\beta_{1}}\cdots\theta^{\alpha_{n}\beta_{n}}\left(\nabla_{\alpha_{1}}\cdots\nabla_{\alpha_{n}}f\right)\left(\nabla_{\beta_{1}}\cdots\nabla_{\beta_{n}}g\right).
\eeq

However since covariant derivatives do not commute, the resulting Moyal product is not associative, $i.e.$ $(f\star g)\star h \neq f\star(g\star h)$. One could consider instead the Kontsevich product \cite{Kontsevich:1997vb}, but its covariant version is also nonassociative \cite{Harikumar:2006xf}. Since one usually implements the noncommutativity through a mapping, the  Seiberg-Witten map \cite{Seiberg:1999vs} up to some order in $\theta^{\mu\nu}$, this procedure usually maintains, up to that order, the associativity, and hence, one chooses the simplest form of covariant deformed product as defined by Eq. (\ref{eq:covmoyal}).

One assumes that for curved space-times that

\beq\label{eq:covtheta}
\nabla_{\alpha}\theta^{\mu\nu} = 0,
\eeq

\noindent $i.e.$ the noncommutative tensor is covariantly constant (see discussion below). This condition was considered in Refs. \cite{Lizzi:2002ib, Harikumar:2006xf} as it generalizes the condition that $\theta^{\mu\nu}$ is constant.

Following Ref. \cite{Bertolami:2002eq}, one consider a scalar field whose commutative analytic potential  $V(\Phi)=\sum_{n=0}^{\infty}\frac{\lambda_{n}}{n!}\Phi^{n}$ is defined by substituting the product between functions by the Moyal product

\beq \label{eq:ncpot}
\tilde{V}(\Phi)=\sum_{n=0}^{\infty}\frac{\lambda_{n}}{n!}\overbrace{\Phi\star\cdots\star\Phi}^{n\;\;times},
\eeq 

\noindent where the tilde denotes a noncommutative function. Although the covariant Moyal product is nonassociative, one can choose an auxiliary condition to keep Eq. (\ref{eq:ncpot}) associative \cite{Bertolami:2002eq} up to second order in the noncommutative parameter\footnote{Another possible way to implement associativity would involve an $m$-dimensional Riemannian manifold with an $SO(m)$ holonomy group and condition (\ref{eq:covtheta}). For a nondegenerate noncommutative parameter, this would require obtaining a ``Maxwell" field without sources in a K\"{a}hler manifold, which for $m$ even, could yield an associative Moyal product. The authors thank Luis Alvarez-Gaum\'{e} for this remark.}

\beq \label{eq:asscond}
\theta^{\mu\nu}\nabla_{\nu}\Phi=0,
\eeq

\noindent and in this case, one can expand the noncommutative potential  (\ref{eq:ncpot}) up to the second order in $\theta^{\mu\nu}$ as \cite{Bertolami:2002eq}:

\beq \label{eq:ncpotexp}
\tilde{V}(\Phi)= V(\Phi) +\frac{1}{2}\frac{d^{2}V(\Phi)}{d\Phi^{2}}\left(-\frac{1}{8}\theta^{\alpha_{1}\beta_{1}}\theta^{\alpha_{2}\beta_{2}}\nabla_{\alpha_{1}}\nabla_{\alpha_{2}}\Phi \nabla_{\beta_{1}}\nabla_{\beta_{2}}\Phi \right).
\eeq

\noindent Eq. (\ref{eq:asscond}) admits two classes of solutions. For  $\det\theta^{\mu\nu}\neq 0$, that is $\theta$ is invertible and then  $\nabla_{\nu}\Phi=0$, a too strong condition for our problem. For $\det\theta^{\mu\nu}= 0$, then $\nabla_{\mu}\Phi$ can be written as powers of the noncommutative parameter, a solution that does not trivialize our problem (cf. Eq. (\ref {eq:condderphi}) and ensued discussion).

One assumes that the gravity sector of the model is not affected by noncommutativity, therefore the space-time is still described  by the usual Einstein equation with noncommutative sources

\beq \label{eq:eeq}
G_{\mu\nu} = \kappa \tilde{T}_{\mu\nu},
\eeq

\noindent where $\kappa =8\pi G$ and in the case under investigation the noncommutative energy-momentum tensor can be split into a  scalar field and  matter fields contributions: $\tilde{T}_{\mu\nu}=\tilde{T}^{\Phi}_{\mu\nu}+\tilde{T}^{\mbox{M}}_{\mu\nu}$. It is further assumed that matter fields satisfy the dominant energy condition\footnote{Physically this condition states that local energy density is positive, that is for any time-like vector $W^{\mu}$, $T_{\mu\nu}W^{\mu}W^{\nu}\geq 0$, and $T_{\mu\nu}W^{\mu}$ is not a space-like vector  \cite{Hawking:1973uf}.}. The noncommutative action then reads

\beq
\tilde{S}=\int d^{4}x\sqrt{-g}\left[\frac{1}{2}g^{\mu\nu}\nabla_{\mu}\Phi\star\nabla_{\nu}\Phi - \tilde{V}(\Phi) + \tilde{\mathcal{L}}_{\mbox{M}} \right]. 
\eeq

The noncommutative generalization of the energy-momentum tensors are given by

\beqa
\tilde{T}^{\mbox{M}}_{\mu\nu} &=& \frac{2}{\sqrt{-g}}\frac{\delta\left(\sqrt{-g}\tilde{\mathcal{L}}_{\mbox{M}}\right)}{\delta g^{\mu\nu}}, \\
\tilde{T}^{\Phi}_{\mu\nu} &=& \frac{1}{2}\left(\nabla_{\mu}\Phi\star\nabla_{\nu}\Phi + \nabla_{\nu}\Phi\star\nabla_{\mu}\Phi\right) -\frac{1}{2} g_{\mu\nu}\nabla_{\rho}\Phi\star\nabla^{\rho}\Phi + g_{\mu\nu}\tilde{V}(\Phi). \label{eq:ncemtsf}
\eeqa

In order to discuss the stability conditions for the scalar field one considers the energy-momentum density for the gravitational field so that the  associated four-momentum vector $p_{\mu}$ for a asymptotically flat space can be written as \cite{Nester:1982tr}

\beq\label{eq:wi}
16\pi G p_{\mu}V^{\mu} = \frac{1}{2}\oint_{S=\partial\Sigma} E^{\sigma\alpha}dS_{\sigma\alpha} = \int_{\Sigma}\nabla_{\alpha}E^{\sigma\alpha}d\Sigma_{\sigma},
\eeq

\noindent where $V^{\mu}=\overline{\epsilon_{0}}\gamma^{\mu}\epsilon_{0}$, $\epsilon_{0}$ represents a constant Dirac spinor, $\Sigma$ is an arbitrary three-dimensional hypersurface and $S$ its boundary  $\partial\Sigma$ at infinity. The two-form $E^{\sigma\alpha}$ is defined as\footnote{Our conventions are the following: the metric signature is $(+,-,-,-)$, $\overline{\epsilon}=\epsilon^{\dagger}\gamma^{0}$, $\{\gamma^{\mu}$,$\gamma^{\nu}\}=2g^{\mu\nu}$, $\sigma^{\mu\nu}=\frac{1}{4}[\gamma^{\mu},\gamma^{\nu}]$, $\epsilon_{0123}=+1$, $\nabla_{\alpha}\epsilon=\partial_{\alpha}\epsilon - \frac{1}{2}\omega^{\mu\nu}_{\alpha}\sigma_{\mu\nu}\epsilon$, $\Gamma^{\sigma\alpha\beta}=\gamma^{[\sigma}
\gamma^{\alpha}\gamma^{\beta]}$, $\Gamma^{\sigma\alpha}=\gamma^{[\sigma}
\gamma^{\alpha]}$.}  

\beq \label{eq:ntf}
E^{\sigma\alpha}=2\left(\overline{\epsilon}\Gamma^{\sigma \alpha\beta}\nabla_{\beta}\epsilon - \overline{\nabla_{\beta}\epsilon}\Gamma^{\sigma \alpha\beta}\epsilon\right),
\eeq

\noindent where $\epsilon$ is a Dirac spinor that at infinity behaves as $\epsilon \rightarrow \epsilon_{0} + \mathcal{O}\left(\frac{1}{r}\right)$. The total energy-momentum can be written with the use of spinor fields. Since one assumes that  gravity is not affected by noncommutativity, the product between spinor fields and gamma matrices is actually the usual one. One further assumes that  spinor fields commute with the noncommutative scalar field. 


\section{Generalized positive energy theorem}\label{sec:gpet}

For supersymmetric theories the method used in Ref. \cite{Nester:1982tr} can be generalized  by replacing Eq. (\ref{eq:ntf}) by

\beq \label{eq:cntf}
\hat{E}^{\sigma\alpha}=2\left(\overline{\epsilon^{i}}\Gamma^{\sigma \alpha\beta}\hat{\nabla}_{\beta}\epsilon^{i} - \overline{\hat{\nabla}_{\beta}\epsilon^{i}}\Gamma^{\sigma \alpha\beta}\epsilon^{i}\right),
\eeq

\noindent where $\hat{\nabla}_{\mu}$ is the supercovariant derivative related to the change of the gravitino field $\psi^{i}_{\;\;\mu}$ under a supersymmetric transformation and $i=1,\ldots,N$ is the number of supersymmetries. One can show that Eq. (\ref{eq:wi}) is then generalized to

\beq \label{eq:gwi}
16\pi G p_{\mu} \overline{\epsilon_{0}^{i}}\gamma^{\mu}\epsilon_{0}^{i} = \int_{\sigma}\left[16\pi G T^{\mbox{M}\sigma}_{\;\;\;\alpha}\overline{\epsilon^{i}}\gamma^{\alpha}\epsilon^{i}  + 4\overline{\hat{\nabla}_{\alpha}\epsilon^{i}}\Gamma^{\sigma \alpha\beta}\hat{\nabla}_{\beta}\epsilon^{i} + \overline{\delta\chi^{a}}\gamma^{\sigma}\delta\chi^{a}   \right]d\Sigma_{\sigma},
\eeq

\noindent where $\delta\chi^{a}$ represents the change of spin-$\frac{1}{2}$ fields under a supersymmetric transformation. In the case of asymptotic Anti-de Sitter (AdS) space-time one requires another term on the L.H.S.  of this equation in order to fix the four-momentum vector $p_{\mu}$. If $T^{\mbox{M}\sigma}_{\;\;\;\alpha}$ satisfies the dominant energy condition, then since  vector $\overline{\epsilon_{0}^{i}}\gamma^{\alpha}\epsilon_{0}^{i}$ is non-space-like the first term in the integrand of Eq.(\ref{eq:gwi}) is positive. Considering the time direction orthogonal to $\Sigma$, thus the last two terms of the R.H.S. of Eq. (\ref{eq:gwi}) can be expressed as\footnote{Latin indices span over $1,2,3$.}

\begin{eqnarray}
4\overline{\hat{\nabla}_{m}\epsilon^{i}}(\gamma^{0}\sigma^{mn}&+&\sigma^{mn}\gamma^{0})\hat{\nabla}_{n}\epsilon^{i} + \left(\delta\chi^{a}\right)^{\dagger}\delta\chi^{a} = \nonumber \\
 &=& -4g^{mn}\left(\hat{\nabla}_{m}\epsilon^{i}\right)^{\dagger}\hat{\nabla}_{n}\epsilon^{i} + 4\left(\hat{\nabla}_{m}\epsilon^{i}\right)^{\dagger}\gamma^{m}\gamma^{n}\hat{\nabla}_{n}\epsilon^{i} + \left(\delta\chi^{a}\right)^{\dagger}\delta\chi^{a}.
\end{eqnarray}

This term is positive definite if one chooses the Witten condition \cite{Gibbons:1983aq}

\beq\label{eq:witten}
\gamma^{n}\hat{\nabla}_{n}\epsilon^{i} = 0.
\eeq

For supersymmetric theories the values of $\hat{\nabla}_{n}\epsilon^{i}$ and $\delta\chi^{a}$ are automatically set by supersymmetry \cite{Gibbons:1983aq,Boucher:1984yx}. If a theory does not admit a supersymmetric extension this setup can be used as discussed in Ref. \cite{Boucher:1984yx}. 

For the scalar field, we define, generalizing the result of Ref. \cite{Bertolami:1987wj},

\begin{eqnarray}
\hat{\nabla}_{\mu}\epsilon^{i} &=& \nabla_{\mu}\epsilon^{i} + \frac{i}{2}\kappa \gamma_{\mu}\tilde{f}^{ij}(\Phi)\epsilon^{j}, \label{eq:defscd} \\
\delta\chi^{a} &=& i\gamma^{\mu}\nabla_{\mu}\Phi\star\tilde{f}^{ai}_{2}(\Phi)\epsilon^{i}+\tilde{f}^{ai}_{3}(\Phi)\epsilon^{i}, \label{eq:defchi}
\end{eqnarray}

\noindent where $\tilde{f}^{ij}(\Phi)$, $\tilde{f}^{ai}_{2}(\Phi)$ and $\tilde{f}^{ai}_{3}(\Phi)$ are noncommutative real scalar functions to be determined. Using the spinor identity $[\nabla_{\mu},\nabla_{\nu}]\epsilon=\frac{1}{2}R^{\alpha\beta}_{\;\;\mu\nu}\sigma_{\alpha\beta}\epsilon$ and Eqs. (\ref{eq:eeq}) and (\ref{eq:ncemtsf}), we can obtain $\nabla_{\alpha}\hat{E}^{\sigma\alpha}$

\begin{eqnarray}\label{eq:ncdiv}
\nabla_{\alpha}\hat{E}^{\sigma\alpha} &=& 2\kappa\tilde{T}^{\mbox{M}\sigma}_{\;\;\;\alpha}\overline{\epsilon^{i}}\gamma^{\alpha}\epsilon^{i}  + 4\overline{\hat{\nabla}_{\alpha}\epsilon^{i}}\star\Gamma^{\sigma \alpha\beta}\hat{\nabla}_{\beta}\epsilon^{i} + \overline{\delta\chi^{a}}\star\gamma^{\sigma}\delta\chi^{a}  \nonumber \\
&+& \left(\tilde{f}^{ai}_{2}(\Phi)\star\nabla_{\alpha}\Phi\right)\star\left(\nabla_{\beta}\Phi\star\tilde{f}^{aj}_{2}(\Phi)\right)\overline{\epsilon}^{i}\Gamma^{\sigma \alpha\beta}\epsilon^{j} \nonumber  \\
&+&\left\{2\kappa\delta^{ij}\left[\frac{\nabla^{\sigma}\Phi\star\nabla_{\alpha}\Phi + \nabla_{\alpha}\Phi\star\nabla^{\sigma}\Phi}{2} - \frac{\delta^{\sigma}_{\;\;\alpha}\nabla^{\rho}\Phi\star\nabla_{\rho}\Phi}{2} \right] \right. \nonumber \\
&-&\left[\left(\tilde{f}^{ai}_{2}(\Phi)\star\nabla_{\alpha}\Phi\right)\star\left(\nabla^{\sigma}\Phi\star\tilde{f}^{aj}_{2}(\Phi)\right) + \left(\tilde{f}^{ai}_{2}(\Phi)\star\nabla^{\sigma}\Phi\right)\star\left(\nabla_{\alpha}\Phi\star\tilde{f}^{aj}_{2}(\Phi)\right) \right. \nonumber \\
&-&\left. \left. \delta^{\sigma}_{\;\;\alpha}\left(\tilde{f}^{ai}_{2}(\Phi)\star\nabla^{\rho}\Phi\right)\star\left(\nabla_{\rho}\Phi\star\tilde{f}^{aj}_{2}(\Phi)\right)\right]\right\}\overline{\epsilon}^{i}\gamma^{\alpha}\epsilon^{j} + i\left[4\kappa\nabla_{\alpha}\tilde{f}^{ij}(\Phi)  \right. \nonumber \\
&-& \left. \left(\tilde{f}^{ai}_{2}(\Phi)\star\nabla_{\alpha}\Phi\right)\star\tilde{f}^{aj}_{3}(\Phi) - \tilde{f}^{ai}_{3}(\Phi)\star\left(\nabla_{\alpha}\Phi\star\tilde{f}^{aj}_{2}(\Phi)\right)\right]\overline{\epsilon}^{i}\Gamma^{\sigma\alpha}\epsilon^{j} \nonumber \\
&+&\left[-\tilde{f}^{ai}_{3}(\Phi)\star\tilde{f}^{aj}_{3}(\Phi)+2\kappa\delta^{ij}\tilde{V}(\Phi)+ 6\kappa^{2}\tilde{f}^{il}(\Phi)\star\tilde{f}^{lj}(\Phi)\right]\overline{\epsilon}^{i}\gamma^{\sigma}\epsilon^{j}  \nonumber 	\\
&+&i\left[\left(\tilde{f}^{ai}_{2}(\Phi)\star\nabla^{\sigma}\Phi\right)\star\tilde{f}^{aj}_{3}(\Phi) - \tilde{f}^{ai}_{3}(\Phi)\star\left(\nabla^{\sigma}\Phi\star\tilde{f}^{aj}_{2}(\Phi)\right) \right]\overline{\epsilon}^{i}\epsilon^{j}.
\end{eqnarray}

One is now in conditions to examine the stability conditions for a noncommutative scalar field. Following Ref. \cite{Boucher:1984yx}, the stability problem consists in obtaining the noncommutative functions $\tilde{f}^{ij}(\Phi)$, $\tilde{f}^{ai}_{2}(\Phi)$ and $\tilde{f}^{ai}_{3}(\Phi)$ for a given $\tilde{V}(\Phi)$ that 
 ensure that Eq. (\ref{eq:ncdiv}) is positive definite.


\section{Stability conditions}\label{sec:stability}

In order to obtain the stability conditions, one must identify Eq. (\ref{eq:ncdiv}) with Eq. (\ref{eq:gwi}). Therefore the coefficients of the last five terms in Eq. (\ref{eq:ncdiv}) must vanish. One first notices that the resulting system of equations is quite difficult to solve, so one assumes that the conditions for indices $i,j,a$ are single valued. This simplifies considerably the system of equations.

One needs now to examine each term at the R. H. S. of Eq. (\ref{eq:ncdiv}). The first term is positive definite given that the matter fields satisfy the dominant energy condition. Choosing ``0" as the direction orthogonal to $\Sigma$, through Eq. (\ref{eq:witten}) one gets that the second and the third terms can be written as

\beq
-4g^{mn}\left(\hat{\nabla}_{m}\epsilon\right)^{\dagger}\star\hat{\nabla}_{n}\epsilon + \left(\delta\chi\right)^{\dagger}\star\delta\chi.
\eeq 

As $\theta^{\mu\nu}$ is covariantly constant, this will be positive definite if one chooses the conditions:

\beqa
\theta^{\mu\nu}\nabla_{\nu}\hat{\nabla}_{n}\epsilon &=&0, \label{eq:condspin} \\
\theta^{\mu\nu}\nabla_{\nu}\delta\chi &=&0. \label{eq:condchi}
\eeqa

One considers now the expansion of a noncommutative function $\tilde{h}(\Phi)$ up to second order in the noncommutative parameter

\beq\label{eq:expncf}
\tilde{h}(\Phi)=h+i\theta^{\mu\nu}h_{\mu\nu}+\theta^{\alpha_{1}\beta_{1}}\theta^{\alpha_{2}\beta_{2}}h_{\alpha_{1}\alpha_{2}\beta_{1}\beta_{2}},
\eeq

\noindent where $h$ is a function of $\Phi$, $h_{\mu\nu}$ is an antisymmetric function of $\Phi$ and its derivatives, and so on. One uses this expansion to compute terms at Eq. (\ref{eq:ncdiv}) that are functions of $\Phi$. 

One looks now to the term proportional to $\overline{\epsilon}\Gamma^{\sigma\alpha\beta}\epsilon$. After using that $\Gamma^{\sigma\alpha\beta}$ is totally antisymmetric and Eq. (\ref{eq:id1}) found in of the Appendix one obtains

\beq
\frac{i\theta^{\mu\nu}}{2}f_{2}^{2}\nabla_{\mu}\nabla_{\alpha}\Phi\nabla_{\nu}\nabla_{\beta}\Phi\overline{\epsilon}\Gamma^{\sigma\alpha\beta}\epsilon-\theta^{\alpha_{1}\beta_{1}}\theta^{\alpha_{2}\beta_{2}}f_{2}f_{2\;\alpha_{2}\beta_{2}}\nabla_{\alpha_{1}}\nabla_{\alpha}\Phi\nabla_{\beta_{1}}\nabla_{\beta}\Phi\overline{\epsilon}\Gamma^{\sigma\alpha\beta}\epsilon,
\eeq

\noindent which vanishes if one chooses that

\beq\label{eq:condderphi}
\theta^{\mu\nu}\nabla_{\mu}\nabla_{\alpha}\Phi =0.
\eeq

Using Eqs. (\ref{eq:id2}) and (\ref{eq:id3}) in the Appendix, the term proportional to $\overline{\epsilon}\epsilon$ can be computed:

\beq
i\theta^{\alpha_{1}\beta_{1}}\theta^{\alpha_{2}\beta_{2}}\left(f_{2}\nabla_{\alpha_{1}}f_{3\;\alpha_{2}\beta_{2}} - f_{3}\nabla_{\alpha_{1}}f_{2\;\alpha_{2}\beta_{2}}  \right)\nabla_{\beta_{1}}\nabla^{\sigma}\Phi\overline{\epsilon}\epsilon,
\eeq

\noindent which vanishes given condition  (\ref{eq:condderphi}). 

The term proportional to $\overline{\epsilon}\gamma^{\alpha}\epsilon$ reads, after using  Eqs. (\ref{eq:condderphi}) and (\ref{eq:id1})

\beqa
& & \left\{ \left(2\kappa -2f_{2}^{2}\right)-i\theta^{\mu\nu}\left(4f_{2}f_{2\;\mu\nu}\right)+\theta^{\alpha_{1}\beta_{1}}\theta^{\alpha_{2}\beta_{2}}	\left(2f_{2\;\alpha_{1}\beta_{1}}f_{2\;\alpha_{2}\beta_{2}}-4f_{2}f_{2\;\alpha_{1}\alpha_{2}\beta_{1}\beta_{2}}\right)\right\} \times \nonumber \\
& & \times \left(\nabla^{\sigma}\Phi\nabla_{\alpha}\Phi - \frac{\delta^{\sigma}_{\;\;\alpha}}{2}\nabla^{\rho}\Phi\nabla_{\rho}\Phi\right)\overline{\epsilon}\gamma^{\alpha}\epsilon.
\eeqa

Clearly, since coefficients of every order in the noncommutative parameter must vanish, one gets

\beq\label{eq:f2}
f_{2}=\sqrt{\kappa}\;\;\;\;\;\;f_{2\;\mu\nu}=0\;\;\;\;\;\;f_{2\;\alpha_{1}\alpha_{2}\beta_{1}\beta_{2}}=0,
\eeq

\noindent and thus that $\tilde{f}_{2}(\Phi)=\sqrt{\kappa}$. 

The term proportional to $\overline{\epsilon}\gamma^{\sigma}\epsilon$ reads after using Eqs. (\ref{eq:asscond}), (\ref{eq:ncpotexp}), (\ref{eq:condderphi}) and (\ref{eq:id4}) 

\beqa
& &\left\{-f_{3}^{2}+2\kappa V(\Phi) + 6\kappa^{2}f^{2} + 2i\theta^{\mu\nu}\left(-f_{3} f_{3\;\mu\nu}+6\kappa^{2}f f_{\mu\nu}\right) + \theta^{\alpha_{1}\beta_{1}}\theta^{\alpha_{2}\beta_{2}}\left(-2f_{3}f_{3\;\alpha_{1}\alpha_{2}\beta_{1}\beta_{2}} \right. \right. \nonumber \\
& &\left. \left. +f_{3\;\alpha_{1}\beta_{1}}f_{3\;\alpha_{2}\beta_{2}} +12\kappa^{2}ff_{\alpha_{1}\alpha_{2}\beta_{1}\beta_{2}} -6\kappa^{2}f_{\alpha_{1}\beta_{1}}f_{\alpha_{2}\beta_{2}}\right) \right\} \overline{\epsilon}\gamma^{\sigma}\epsilon. \label{eq:gammaalpha}
\eeqa

\noindent In order to proceed one assumes that $\tilde{f}(\Phi)=a+b\Phi\star\Phi$, where constants $a$ and $b$ must be obtained by the boundary conditions of the system of equations; this condition generalizes the procedure of Ref. \cite{Bertolami:1987wj}. Using Eq. (\ref{eq:condderphi}) one gets

\beqa
-f_{3}^{2}+2\kappa V(\Phi) + 6\kappa^{2}f^{2} &=&0, \label{eq:f3b} \\
f_{3}f_{3\;\mu\nu}&=&0, \label{eq:f3final} \\
f_{3\;\alpha_{1}\alpha_{2}\beta_{1}\beta_{2}}&=&\frac{f_{3\;\alpha_{1}\beta_{1}}f_{3\;\alpha_{2}\beta_{2}}}{2f_{3}}. \label{eq:f3a}
\eeqa 

\noindent Eq. (\ref{eq:f3final}) yields

\beq\label{eq:f3munu}
f_{3\;\mu\nu}=0,
\eeq

\noindent substituting this into Eq. (\ref{eq:f3a}), it follows that

\beq\label{eq:f3so}
f_{3\;\alpha_{1}\alpha_{2}\beta_{1}\beta_{2}}=0.
\eeq

Finally, the term proportional to $\overline{\epsilon}\Gamma^{\sigma\alpha}\epsilon$ is given by

\beqa
& &i\left\{
\left[4\kappa\left(\frac{df}{d\Phi}\right)-2f_{2}f_{3}\right]\nabla_{\alpha}\Phi-2i\theta^{\mu\nu}\left(f_{2}f_{3\;\mu\nu}+f_{3}f_{2\;\mu\nu}\right) \right. \nonumber \\
& & \left. - 2\theta^{\alpha_{1}\beta_{1}}\theta^{\alpha_{2}\beta_{2}}
\left(f_{3}f_{2\;\alpha_{1}\alpha_{2}\beta_{1}\beta_{2}}+f_{2}f_{3\;\alpha_{1}\alpha_{2}\beta_{1}\beta_{2}}-f_{2\;\alpha_{1}\beta_{1}}f_{3\;\alpha_{2}\beta_{2}} \right)\right\} \overline{\epsilon}\Gamma^{\sigma\alpha}\epsilon. 
\eeqa

Using Eqs. (\ref{eq:f2}), (\ref{eq:f3munu}) and (\ref{eq:f3so}), it yields

\beq
4\kappa\left(\frac{df}{d\Phi}\right)-2f_{2}f_{3} =0. \label{eq:derphif} 
\eeq

Thus, the problem of stability consists in solving the system of equations

\beqa
2\sqrt{\kappa}\left(\frac{df}{d\Phi}\right)&=&f_{3}, \label{eq:stab1} \\
-f_{3}^{2}+2\kappa V(\Phi) + 6\kappa^{2}f^{2} &=&0. \label{eq:stab2}
\eeqa

However, this is precisely the set of equations for the commutative case for a quartic potential solved in  Ref. \cite{Bertolami:1987wj}. Our result is then  that the stability conditions for a scalar with a noncommutative potential are not affected by noncommutativity.

Let us now examine the consistency of Eqs. (\ref{eq:condspin}) and (\ref{eq:condchi}) after solving the stability conditions  (\ref{eq:f2}), (\ref{eq:f3munu}) and (\ref{eq:f3so}). At first order in perturbation of the noncommutative parameter, one obtains

\beq
\theta^{\mu\nu}\nabla_{\nu}\hat{\nabla}_{n}\epsilon=\theta^{\mu\nu}\nabla_{\nu}\nabla_{n}\epsilon+\frac{i\kappa}{2}\gamma_{n}f(\Phi)\theta^{\mu\nu}\nabla_{\nu}\epsilon	+\frac{i\kappa}{2}\gamma_{n}\left(\theta^{\mu\nu}\nabla_{\nu}f(\Phi)\right)\epsilon=0.
\eeq

\noindent The first two terms vanish by the assumption that spinors are not affected by noncommutativity. The last term vanishes on account of Eq. (\ref{eq:asscond}). Therefore, this equation is consistent with the results obtained above. One can also show that $\theta^{\mu\nu}\nabla_{\nu}\delta\chi=0$, using the assumption that spinors are not altered by noncommutativity and Eqs. (\ref{eq:asscond}) and (\ref{eq:condderphi}).


\section{Space-time with horizons}\label{sec:bhspt}

One considers now space-time configurations which admit horizons. In this situation, the divergence theorem must be modified so to include the horizon

\beq\label{eq:sggt}
\frac{1}{2}\oint_{S} \hat{E}^{\sigma\alpha}dS_{\sigma\alpha} - \frac{1}{2}\oint_{H} \hat{E}^{\sigma\alpha}dS_{\sigma\alpha} = \int_{\Sigma}\nabla_{\alpha}\hat{E}^{\sigma\alpha}d\Sigma_{\sigma},
\eeq

\noindent where $H$ denotes the horizon. Clearly, if the second term in the L. H. S. of Eq. (\ref{eq:sggt}) vanishes the presence of horizons does not affect the stability conditions obtained in Section \ref{sec:stability}. 

Following Ref. \cite{Straumann:1984xf} one introduces a orthonormal tetrad field at the horizon $\{e_{\hat{\mu}}\}$, where $e_{\hat{0}}$ is normal to the hypersurface $\Sigma$, $e_{\hat{1}}$ is normal to the two-surface $H$ and $e_{\hat{A}}$ ($A=2,3$) are tangent to $H$. Using this coordinate
system then one  has only to evaluate the term   

\beq
\oint_{H} \hat{E}^{\hat{0}\hat{1}}dS_{\hat{0}\hat{1}}.
\eeq

For simplicity one omits the hat on the indices. First one restricts the two-form to $\Sigma$, and thus through  Witten's condition $\gamma^{a}\hat{\nabla}_{a}\epsilon=0$, one finds that

\beq
\left.\hat{E}^{0a}\right|_{\Sigma} = -2\epsilon^{\dagger}\hat{\nabla}^{a}\epsilon + \mbox{h. c.}\;\;\;\;.
\eeq
 
\noindent Using the definition of the supercovariant derivative and $\nabla_{b}\epsilon= ^{(3)}\!\!\nabla_{b}\epsilon+\frac{1}{2}K_{ab}\gamma^{0}\gamma^{a}\epsilon$, where $^{(3)}\!\nabla_{b}$ is the intrinsic three-dimensional covariant derivative and $K_{ab}$ is the second fundamental form of $\Sigma$, then the value of  two-form on $H$ is given by

\begin{eqnarray}
\left.\hat{E}^{01}\right|_{H} &=& 2\epsilon^{\dagger}\hat{\nabla}_{1}\epsilon + \mbox{h. c.} \nonumber \\
&=& 2\epsilon^{\dagger} { }^{(3)}\!\nabla_{1}\epsilon + K_{1b}\epsilon^{\dagger}\gamma^{0}\gamma^{b}\epsilon - i\kappa\tilde{f}(\Phi)\epsilon^{\dagger}\gamma^{1}\epsilon + \mbox{h. c.} \label{eq:e01h1}
\end{eqnarray}

From Witten's condition: 

\beq\label{eq:nabla1}
{ }^{(3)}\!\nabla_{1}\epsilon = \gamma^{1}\gamma^{A}{ }^{(3)}\!\nabla_{A}\epsilon - \frac{1}{2}K\gamma^{1}\gamma^{0}\epsilon + \frac{3}{2}i\kappa \tilde{f}(\Phi)\gamma^{1}\epsilon,
\eeq

\noindent where $K=K^{a}_{\;\;a}$. Substituting Eq. (\ref{eq:nabla1}) into Eq. (\ref{eq:e01h1}) and using that ${ }^{(3)}\!\nabla_{A}\epsilon={ }^{(2)}\!\nabla_{A}\epsilon - \frac{1}{2}J_{AB}\gamma^{1}\gamma^{B}\epsilon$\footnote{${ }^{(2)}\!\nabla_{A}$ is the intrinsic covariant derivative on $H$ and $J_{AB}$ is the second fundamental form on $H$ with $J=J^{A}_{\;\;A}$.}, it follows that

\beq\label{eq:e01h2}
\left.\hat{E}^{01}\right|_{H}=\epsilon^{\dagger}\left[2\gamma^{1}\gamma^{A}\mathcal{D}_{A} - \left(J + \left(K + K_{11}\right)\gamma^{1}\gamma^{0}\right) + 2i\kappa \tilde{f}(\Phi)\gamma^{1} \right]\epsilon + \mbox{h. c.}\;\;\;\;,
\eeq 

\noindent where $\mathcal{D}_{A} \equiv \left({ }^{(2)}\!\nabla_{A} - \frac{1}{2}K_{1A}\gamma^{1}\gamma^{0} \right)$. A further condition is required to restrict the spinor field on $H$. This has been put forward in Ref. \cite{Gibbons:1982jg}, namely: $\gamma^{1}\gamma^{0}\epsilon =\epsilon$. Now Eq. (\ref{eq:e01h2}) reads

\beq\label{eq:e01h3}
\left.\hat{E}^{01}\right|_{H}=\epsilon^{\dagger}\left[2\gamma^{1}\gamma^{A}\mathcal{D}_{A} - \left(J + K + K_{11}\right)\right]\epsilon + 2i\kappa \tilde{f}(\Phi)\epsilon^{\dagger}\gamma^{1}\epsilon + \mbox{h. c.} \;\;\;\; .
\eeq 

Notice that $\left(J + K + K_{11}\right)=-\sqrt{2}\psi$, where $\psi$ is the expansion scalar \cite{Straumann:1984xf}, which is related to the rate of increase of the absolute value of the element of area. If two neighbouring geodesics are
converging, then $\psi<0$, if instead they  diverge, then $\psi>0$.  This quantity vanishes if $H$ is an apparent horizon. Given that $\gamma^{1}\gamma^{0}$ anticommutes with $\gamma^{1}\gamma^{A}\mathcal{D}_{A}$ and with $\gamma^{1}$, then

\beq\label{eq:ac1}
2i\kappa\tilde{f}(\Phi)\epsilon^{\dagger}\gamma^{1}\epsilon=2i\kappa\tilde{f}(\Phi)\epsilon^{\dagger}\gamma^{1}\gamma^{1}\gamma^{0}\epsilon=-2i\kappa\tilde{f}(\Phi)\epsilon^{\dagger}\gamma^{1}\gamma^{0}\gamma^{1}\epsilon=-2i\kappa\tilde{f}(\Phi)\epsilon^{\dagger}\gamma^{1}\epsilon=0,
\eeq
	
\noindent and

\beq\label{eq:ac2}
2\epsilon^{\dagger}\gamma^{1}\gamma^{A}\mathcal{D}_{A}\epsilon=2\epsilon^{\dagger}\gamma^{1}\gamma^{A}\mathcal{D}_{A}\gamma^{1}\gamma^{0}\epsilon=-2\epsilon^{\dagger}\gamma^{1}\gamma^{0}\gamma^{1}\gamma^{A}\mathcal{D}_{A}\epsilon=-2\epsilon^{\dagger}\gamma^{1}\gamma^{A}\mathcal{D}_{A}\epsilon=0.
\eeq

Thus, choosing the boundary $H$ to be an apparent horizon, from Eqs. (\ref{eq:ac1}) and (\ref{eq:ac2}) one finds that 

\beq
\oint_{H} \hat{E}^{\hat{0}\hat{1}}dS_{\hat{0}\hat{1}}=0,
\eeq

\noindent and therefore the presence of spaces with horizons does not affect the stability conditions found in Section \ref{sec:stability}.

\section{Conclusions}\label{sec:conclusions}

In this work the stability conditions for a noncommutative scalar field coupled to gravity have been examined. Gravity is assumed not to be affected by noncommutativity and also that in the Moyal product usual derivatives are replaced by covariant derivatives. Associativity is ensured through  an auxiliary condition, namely $\theta^{\mu\nu}\nabla_{\nu}\Phi=0$. It is then found that for a scalar field with a polynomial potential, the stability conditions are the very ones for the commutative case studied in Ref. \cite{Bertolami:1987wj}.

At first sight one might think that this result was already expected, given that no noncommutative corrections to $\tilde{V}(\Phi)$ and $\tilde{f}(\Phi)$ were considered up to the second order in $\theta$. This is not quite the case as one encounters that we obtain a nontrivial condition  for the term proportional to $\overline{\epsilon}\Gamma^{\sigma\alpha\beta}\epsilon$ (Eq. (\ref{eq:condderphi})), which is actually absent in the commutative case. It is interesting to point out that the obtained conditions for the stability of a noncommutative scalar field, Eqs. (\ref{eq:condspin}), (\ref{eq:condchi}) and (\ref{eq:condderphi}), are structurally related with the associativity condition, Eq. (\ref{eq:asscond}).

Finally, it has also been shown that the contribution of the surface integral $\oint_{H}\hat{E}^{\sigma\alpha}dS_{\sigma\alpha}$ on an apparent horizon vanishes. This means that stability results are not altered whether one considers space-time configurations with an apparent horizon.


\section*{Acknowledgments}


The work of O. B. is partially supported by the Funda\c{c}\~{a}o para a Ci\^{e}ncia e a Tecnologia (FCT) under the project POCI/FIS/56093/2004. The work of C. A. D. Z. is fully supported by the FCT fellowship SFRH/BD/29446/2006.


\appendix

\section{Appendix}
In Section \ref{sec:stability} after expanding terms up to second order in $\theta$, one cannot fail to see the similarity of many of the obtained terms. Here one derives all  terms encountered  in  Eq. (\ref{eq:ncdiv}). One uses the expansion of noncommutative functions in powers of the noncommutative parameter (Eq. (\ref{eq:expncf})), the definition of the covariant Moyal product (Eq. (\ref{eq:covmoyal})), and the associativity condition Eqs. (\ref{eq:covtheta}) and (\ref{eq:asscond}). Four types of noncommutative products are found:

\beqa
\left(\tilde{f}(\Phi)\star\nabla_{\alpha}\Phi\right)&\star&\left(\nabla_{\beta}\Phi\star\tilde{f}(\Phi)\right)=f^{2}\nabla_{\alpha}\Phi\nabla_{\beta}\Phi + i\theta^{\mu\nu}\left[\frac{f^{2}}{2}\nabla_{\mu}\nabla_{\alpha}\Phi\nabla_{\nu}\nabla_{\beta}\Phi + 2ff_{\mu\nu}\nabla_{\alpha}\Phi\nabla_{\beta}\Phi\right]\nonumber \\
&+&\theta^{\alpha_{1}\beta_{1}}\theta^{\alpha_{2}\beta_{2}}\left[-\frac{f^{2}\nabla_{\alpha_{1}}\nabla_{\alpha_{2}}\nabla_{\alpha}\Phi\nabla_{\beta_{1}}\nabla_{\beta_{2}}\nabla_{\beta}\Phi}{8} + 2ff_{\alpha_{1}\alpha_{2}\beta_{1}\beta_{2}}\nabla_{\alpha}\Phi\nabla_{\beta}\Phi\right.	\nonumber \\
&-&\left. ff_{\alpha_{2}\beta_{2}}\nabla_{\alpha_{1}}\nabla_{\alpha}\Phi\nabla_{\beta_{1}}\nabla_{\beta}\Phi - f_{\alpha_{1}\beta_{1}}f_{\alpha_{2}\beta_{2}}\nabla_{\alpha}\Phi\nabla_{\beta}\Phi\right], \label{eq:id1}
\eeqa

\beqa
\left(\tilde{f}(\Phi)\star\nabla^{\sigma}\Phi\right)\star\tilde{g}(\Phi)&=&fg\nabla^{\sigma}\Phi+i\theta^{\mu\nu}\left(fg_{\mu\nu}+gf_{\mu\nu}\right)\nabla^{\sigma}\Phi \nonumber \\
&+&\theta^{\alpha_{1}\beta_{1}}\theta^{\alpha_{2}\beta_{2}}\left[\left(f\nabla_{\alpha_{1}}g_{\alpha_{2}\beta_{2}}-g\nabla_{\alpha_{1}}f_{\alpha_{2}\beta_{2}}\right)\frac{\nabla_{\beta{1}}\nabla^{\sigma}\Phi}{2}\right. \nonumber \\
&+& \left.\left(fg_{\alpha_{1}\alpha_{2}\beta_{1}\beta_{2}}+gf_{\alpha_{1}\alpha_{2}\beta_{1}\beta_{2}}-f_{\alpha_{1}\beta_{1}}g_{\alpha_{2}\beta_{2}}\right)\nabla^{\sigma}\Phi\right], \label{eq:id2}
\eeqa

\beqa
\tilde{g}(\Phi)\star\left(\nabla^{\sigma}\Phi\star\tilde{f}(\Phi)\right)&=&fg\nabla^{\sigma}\Phi+i\theta^{\mu\nu}\left(fg_{\mu\nu}+gf_{\mu\nu}\right)\nabla^{\sigma}\Phi \nonumber \\
&+&\theta^{\alpha_{1}\beta_{1}}\theta^{\alpha_{2}\beta_{2}}\left[\left(-f\nabla_{\alpha_{1}}g_{\alpha_{2}\beta_{2}}+g\nabla_{\alpha_{1}}f_{\alpha_{2}\beta_{2}}\right)\frac{\nabla_{\beta{1}}\nabla^{\sigma}\Phi}{2}\right. \nonumber \\
&+& \left.\left(fg_{\alpha_{1}\alpha_{2}\beta_{1}\beta_{2}}+gf_{\alpha_{1}\alpha_{2}\beta_{1}\beta_{2}}-f_{\alpha_{1}\beta_{1}}g_{\alpha_{2}\beta_{2}}\right)\nabla^{\sigma}\Phi\right], \label{eq:id3}
\eeqa

\beq \label{eq:id4}
\tilde{f}(\Phi)\star\tilde{f}(\Phi)=f^{2}+2i\theta^{\mu\nu}ff_{\mu\nu}+\theta^{\alpha_{1}\beta_{1}}\theta^{\alpha_{2}\beta_{2}}\left(2ff_{\alpha_{1}\alpha_{2}\beta_{1}\beta_{2}}-f_{\alpha_{1}\beta_{1}}f_{\alpha_{2}\beta_{2}}\right).
\eeq

\newpage


\appendix
\newpage
\vfill


\begin{thebibliography}{99}

\bibitem{Connes:1996gi}
  A.~Connes,
  Commun.\ Math.\ Phys.\  {\bf 182}, 155 (1996)
  [arXiv:hep-th/9603053].

\bibitem{Seiberg:1999vs}
  N.~Seiberg and E.~Witten,
  JHEP {\bf 9909}, 032 (1999)
  [arXiv:hep-th/9908142].

\bibitem{Snyder:1947}
  H. ~S. Snyder,
  Phys.\ Rev.\  {\bf 71}            (1947) 38. 

\bibitem{Szabo:2001kg}
  R.~J.~Szabo,
  Phys.\ Rept.\  {\bf 378}, 207 (2003)
  [arXiv:hep-th/0109162].


\bibitem{Douglas:2001ba}
  M.~R.~Douglas and N.~A.~Nekrasov,
  Rev.\ Mod.\ Phys.\  {\bf 73}, 977 (2001)
  [arXiv:hep-th/0106048].


\bibitem{Bertolami:2003nm}
  O.~Bertolami and L.~Guisado,
  JHEP {\bf 0312}, 013 (2003)
  [arXiv:hep-th/0306176].


\bibitem{Lizzi:2002ib}
  F.~Lizzi, G.~Mangano, G.~Miele and M.~Peloso,
  JHEP {\bf 0206}, 049 (2002)
  [arXiv:hep-th/0203099].

\bibitem{Bertolami:2002eq}
  O.~Bertolami and L.~Guisado,
  Phys.\ Rev.\  D {\bf 67}             (2003) 025001 
  [arXiv:gr-qc/0207124].
  

\bibitem{Gamboa:2001fg}
  J.~Gamboa, M.~Loewe, F.~Mendez and J.~C.~Rojas,
  Mod.\ Phys.\ Lett.\  A {\bf 16}, 2075 (2001)
  [arXiv:hep-th/0104224]; J.~Gamboa, M.~Loewe and J.~C.~Rojas,
  Phys.\ Rev.\  D {\bf 64}, 067901 (2001)
  [arXiv:hep-th/0010220]; A.~P.~Balachandran, A.~R.~Queiroz, A.~M.~Marques and P.~Teotonio-Sobrinho,
  Phys.\ Rev.\  D {\bf 77}, 105032 (2008)
  [arXiv:0706.0021 [hep-th]].




\bibitem{Zhang:2004yu}
  J.~Z.~Zhang,
  Phys.\ Rev.\ Lett.\  {\bf 93}, 043002 (2004)
  [arXiv:hep-ph/0405143];  J.~Z.~Zhang,
  Phys.\ Lett.\  B {\bf 584}, 204 (2004)
  [arXiv:hep-th/0405135].


\bibitem{Bertolami:2005jw}
  O.~Bertolami, J.~G.~Rosa, C.~M.~L.~de Arag\~{a}o, P.~Castorina and D.~Zappal\`{a},
  Phys.\ Rev.\  D {\bf 72}, 025010 (2005)
  [arXiv:hep-th/0505064].

\bibitem{Acatrinei:2003id}
  C.~S.~Acatrinei,
  Mod.\ Phys.\ Lett.\  A {\bf 20}, 1437 (2005)
  [arXiv:hep-th/0311134].

\bibitem{Bertolami:2005ud}
  O.~Bertolami, J.~G.~Rosa, C.~M.~L.~de Arag\~{a}o, P.~Castorina and D.~Zappal\`{a},
  Mod.\ Phys.\ Lett.\  A {\bf 21}, 795 (2006)
  [arXiv:hep-th/0509207].

\bibitem{Bastos:2006kj}
  C.~Bastos and O.~Bertolami,
  Phys.\ Lett.\  A {\bf 372}, 5556 (2008)
  [arXiv:gr-qc/0606131].

\bibitem{Bastos:2006ps}
  C.~Bastos, O.~Bertolami, N.~C.~Dias and J.~N.~Prata,
  J.\ Math.\ Phys.\  {\bf 49}, 072101 (2008)
  [arXiv:hep-th/0611257].

\bibitem{GarciaCompean:2001wy}
  H.~Garcia-Compean, O.~Obreg\'{o}n and C.~Ramirez,
  Phys.\ Rev.\ Lett.\  {\bf 88}, 161301 (2002)
  [arXiv:hep-th/0107250].

\bibitem{Barbosa:2004kp}
  G.~D.~Barbosa and N.~Pinto-Neto,
  Phys.\ Rev.\  D {\bf 70}, 103512 (2004)
  [arXiv:hep-th/0407111].

\bibitem{Bastos:2007bg}
  C.~Bastos, O.~Bertolami, N.~C.~Dias and J.~N.~Prata,
  Phys.\ Rev.\  D {\bf 78}, 023516 (2008)
  [arXiv:0712.4122 [gr-qc]].


\bibitem{Witten:1981mf}
  E.~Witten,
  Commun.\ Math.\ Phys.\  {\bf 80}, 381 (1981).


\bibitem{ChoquetBruhat:1985xy}
  Y.~Choquet-Bruhat,
  ``Positive Energy Theorems,''
{\it  In *Les Houches 1983, Proceedings, Relativity, Groups and Topology, Ii*, 739-785}






\bibitem{Gibbons:1982fy}
  G.~W.~Gibbons and C.~M.~Hull,
  Phys.\ Lett.\  B {\bf 109}, 190 (1982).

\bibitem{Gibbons:1983aq}
  G.~W.~Gibbons, C.~M.~Hull and N.~P.~Warner,
  Nucl.\ Phys.\  B {\bf 218}, 173 (1983).





\bibitem{Boucher:1984yx}
  W.~Boucher,
  Nucl.\ Phys.\  B {\bf 242}, 282 (1984).

\bibitem{Bertolami:1987wj}
  O.~Bertolami,
  Phys.\ Lett.\  B {\bf 186}, 161 (1987).




\bibitem{Gibbons:1982jg}
  G.~W.~Gibbons, S.~W.~Hawking, G.~T.~Horowitz and M.~J.~Perry,
  Commun.\ Math.\ Phys.\  {\bf 88}, 295 (1983).

\bibitem{Bayen:1977ha}
  F.~Bayen, M.~Flato, C.~Fronsdal, A.~Lichnerowicz and D.~Sternheimer,
  Annals Phys.\  {\bf 111}, 61 (1978).

\bibitem{BarcelosNeto:2002bh}
  J.~Barcelos-Neto,
  arXiv:hep-th/0212094.


\bibitem{Harikumar:2006xf}
  E.~Harikumar and V.~O.~Rivelles,
  Class.\ Quant.\ Grav.\  {\bf 23}, 7551 (2006)
  [arXiv:hep-th/0607115].
  
\bibitem{Kontsevich:1997vb}
  M.~Kontsevich,
  Lett.\ Math.\ Phys.\  {\bf 66} (2003) 157
  [arXiv:q-alg/9709040].

\bibitem{Hawking:1973uf}
  S.~W.~Hawking and G.~F.~R.~Ellis,
  ``The Large scale structure of space-time,''
{\it  Cambridge University Press, Cambridge, 1973}

\bibitem{Nester:1982tr}
  J.~A.~Nester,
  Phys.\ Lett.\  A {\bf 83}, 241 (1981).

\bibitem{Straumann:1984xf}
  N.~Straumann,
  ``General Relativity And Relativistic Astrophysics,''
{\it  Berlin, Germany: Springer ( 1984) 459 P. ( Texts and Monographs In Physics)}



\end{thebibliography}
\end{document}